\documentclass[a4paper,11pt]{article}
\pdfoutput=1
\usepackage{jinstpub}
\usepackage{lineno}
\title{\boldmath The ABALONE Photosensor}

\author[a,b,1]{V.~D'Andrea,\note{Corresponding author.}}
\author[b]{R. Biondi,}
\author[b,c]{C. Ferrari,}
\author[a,b]{A.D. Ferella,}
\author[d]{J. Mahlstedt,}
\author[b,e]{and G. Pieramico}

\affiliation[a]{Dipartimento di Scienze Fisiche e Chimiche, Università degli Studi dell'Aquila, L'Aquila I-67100, Italy}
\affiliation[b]{INFN, Laboratori Nazionali del Gran Sasso, Assergi, L'Aquila I-67100, Italy}
\affiliation[c]{Gran Sasso Science Institute (GSSI), L'Aquila I-67100, Italy}
\affiliation[c]{Oskar Klein Centre, Department of Physics, Stockholm University, AlbaNova, Stockholm SE-10691, Sweden}
\affiliation[d]{Dipartimento di Fisica, Università La Sapienza di Roma, Roma I-00185, Italy}

\emailAdd{valerio.dandrea@lngs.infn.it}

\abstract{The ABALONE is a new type of photosensor produced by PhotonLab, Inc. with cost effective mass production, robustness and high performance. This modern technology provides sensitivity to visible and UV light, exceptional radio-purity and excellent detection performance in terms of intrinsic gain, afterpulsing rate, timing resolution and single-photon sensitivity. For these reasons, the ABALONE can have many fields of application, including particle physics experiments, such as DARWIN, and medical imaging. This new hybrid photosensor, that works as light intensifier, is based on the acceleration in vacuum of photoelectrons generated in a traditional photosensor cathode and guided towards a window of scintillating material that can be read from the outside through a silicon photomultiplier.
In this work we present the simulation of the ABALONE and the results from operation at room temperature. The goal of the characterization is the evaluation of the gain, the response in time and the single photoelectron spectrum as a function of the electric field and the photoelectron emission angle. Details of future tests will be also discussed.}

\proceeding{LIDINE 2021: LIght Detection In Noble Elements\\
  14-18 September 2021}

\begin{document}

\maketitle
\flushbottom

\section{Introduction}
\label{sec:intro}

Photosensor have undergone great development in recent years, aiming costs reduction and performance improvement. Wide used photosensors are the photomultiplier tubes (PMTs), built through a long and expensive process of combining materials. To have a low-cost mass production, new construction methods and different materials must be used.
In recent years, this achievement has been reached with a new hybrid photosensor concept, called ABALONE \cite{abalone2017,abalone2018}. The ABALONE bypassed critical manufacturing problems avoiding all materials other than glass among the components entering the uninterrupted vacuum production line. A "glass alone" concept provides, unlike metals, zero trapped gases. The following sections describe the main features of the ABALONE (section \ref{Sec:abalone}) and its characterization, including simulation (section \ref{Sec:simulation}) and experimental data analysis (section \ref{Sec:results}).

\section{The ABALONE photosensor}\label{Sec:abalone}
\begin{figure}[htbp]
\centering % \begin{center}/\end{center}
\includegraphics[width=.4\textwidth]{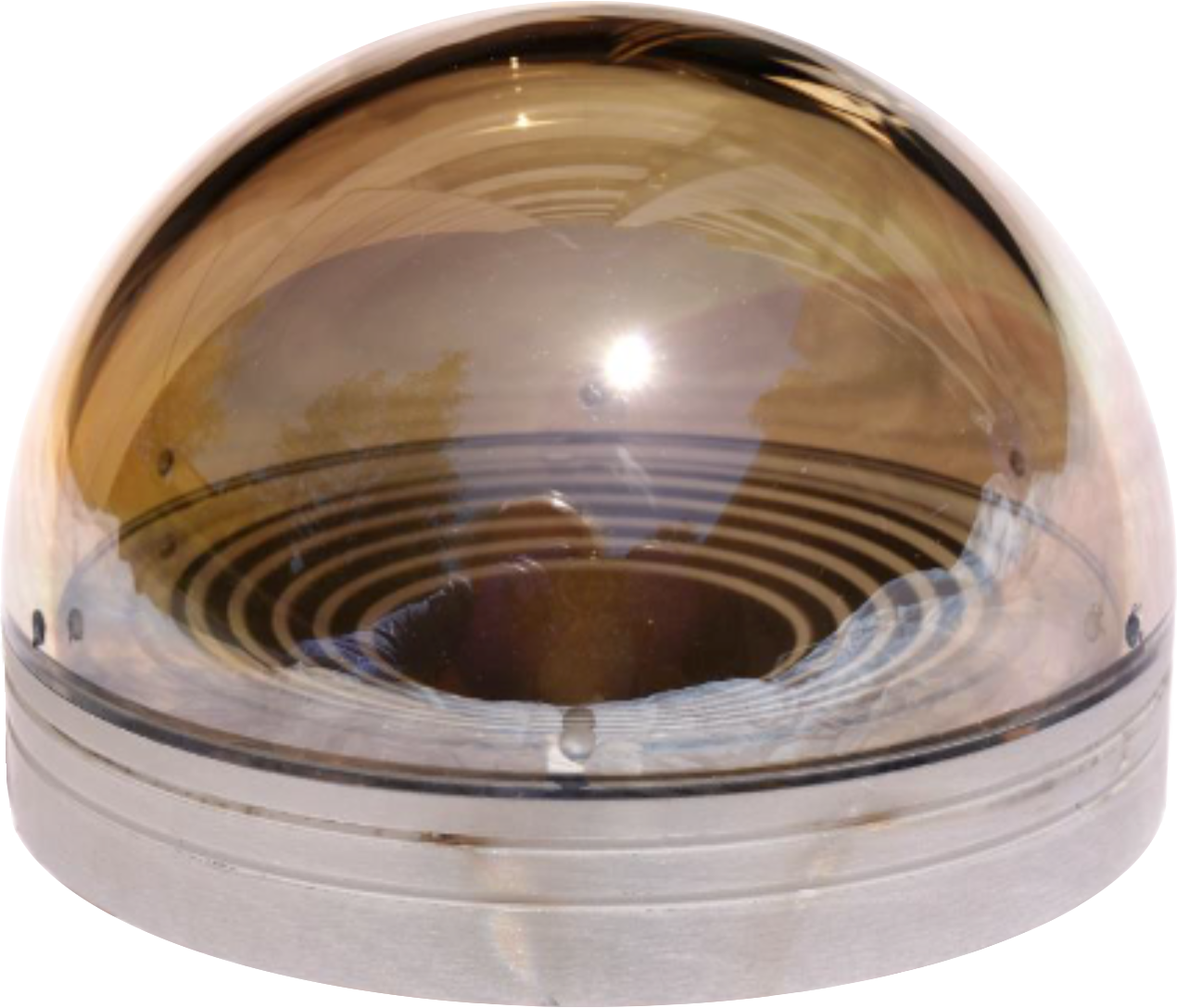}
\qquad
\includegraphics[width=.5\textwidth]{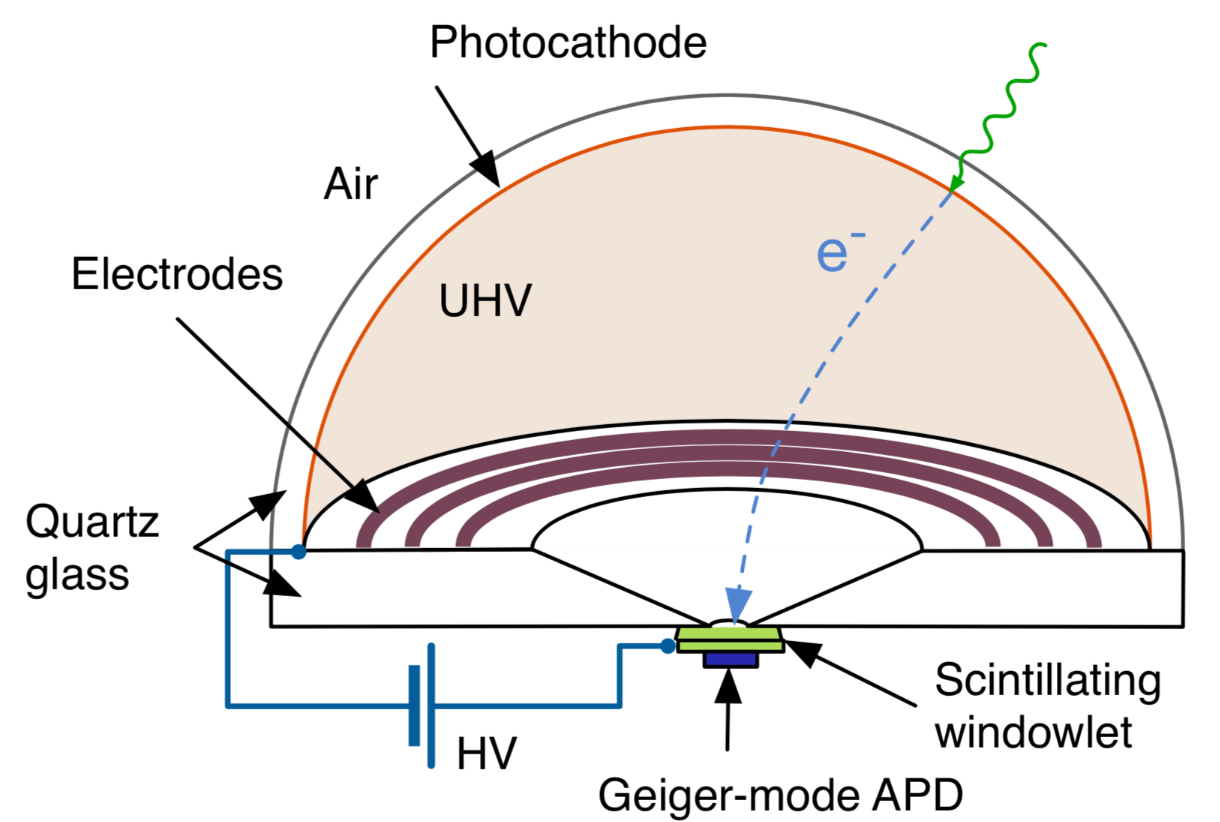}
\caption{\label{Fig:abalone} The ABALONE Photosensor prototype developed for the IceCube extension project \cite{abalone2018} (left) and the ABALONE schematic indicating the photosensor components (right).}
\end{figure}

The ABALONE photosensor design comprises only three industrially prefabricated glass components: the dome, the base plate and the windowlet, made in a continuous production-line process. In figure \ref{Fig:abalone} (left) a picture of an ABALONE is shown.

\subsection{Construction}

The ultra-high vacuum assembly line consists of four chambers: the load lock chamber, the thin-film deposition chamber, the photocathode deposition chamber and assembly chamber. The R\&D project of PhotonLab, Inc. \cite{abalone2017} has recently designed and acquired a production-oriented facility, in order to perform application specific and gradually increase production. One of the strengths of this technology is the thin-film vacuum-sealing technique. Two glass-to-glass bonds are established using the "smart multi-functional thin-film", one between the dome and the base plate, and the other between the base plate and the windowlet. The dome and the base plate are custom-made of fused silica. A thin Cs$_3$Sb alkali-antimonide photocathode lines the inner surface of the dome. The windowlet is made of a lutetium–yttrium oxyorthosilicate (LYSO) scintillator crystal \cite{lyso}, that provides high light output with a light yield of about 30~ph/keV, fast decay time, excellent energy resolution and low cost. On the vacuum side, the windowlet is coated with a chromium-gold layer, which connects to ground potential and acts as a reflector. The ABALONE has an outer dome diameter of 11~cm.

\subsection{Principle of operation}

The ABALONE works as a light intensifier. The photocathode in the inner surface of the dome converts photons in electrons via photoelectric effect. Photoelectrons (PEs) are released in the internal ABALONE volume in vacuum and from here accelerated towards the windowlet by an externally electrostatic field. At this purpose a high voltage generator ($\sim 25$~kV) is connected to the base plate. The glass-to-glass bonding method allows the high voltage to pass trough the base plate and reach the photocathode, in the same way the ground potential is connected to the windowlet. Following the electrical field, the PEs are focused into the 3~mm inner surface of the windowlet. Before reaching the scintillator, PEs first penetrate through the chromium-gold layer suffering a certain loss of energy (about 0.3~keV for kinetic energy of 20~keV \cite{abalone2017}). PEs inside the scintillator, with the exception of those that back-scatter, lose all the energy, part of which is used by the crystal to generate secondary photons. An avalanche photodiode has to be optically coupled to the external surface of the windowlet to read these secondary photons. Around 35\% of PEs are back scatter from the scintillator into vacuum after having lost part of energy. These can either return back in the scintillator depositing their entire kinetic energy or, if they have left the LYSO crystal with a small fraction of their initial energy, they may be deflected significantly not returning back to the windowlet. In this case, PEs deposit only a fraction of their energy in the scintillator, thus number of secondary photons is not proportional to the initial kinetic energy. This effect has been addressed and quantified in existent ABALONE prototypes. One of the objectives of the R\&D activity (including the work described here) is the reduction of the fraction of non-returning PEs.
A typical ABALONE configuration consists of a high voltage ($V$) of 25~kV, thus PEs have 25~keV kinetic energy. A complete energy release in LYSO produces $\sim750$ photons. If the light collection is 10\%, $\sim 75$ are detected. With a SiPM gain of $\sim 10^6$, the total gain of system ABALONE and SiPM is $\sim 10^8$.

\section{ABALONE simulation}\label{Sec:simulation}

\begin{figure}[htbp]
\centering
\includegraphics[width=0.5\textwidth]{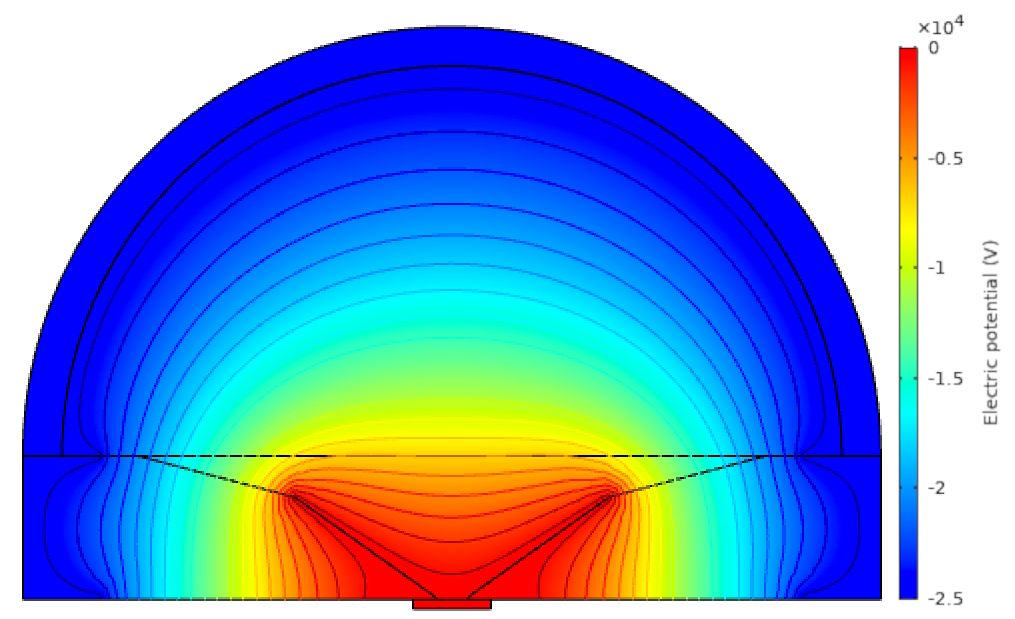}
\caption{\label{fig:PotTrack} ABALONE electrical field lines with $V=25$~kV obtained with the COMSOL Multiphysics software.}
\end{figure}

The full photosensor simulation, including reconstruction of ABALONE geometry, PE trajectories and SiPM light response is performed with the COMSOL Multiphysics \cite{comsol} and GEANT4 \cite{geant4} toolkits. More details on this work can be found in \cite{giulia}. The electrical potential within the ABALONE is reproduced by COMSOL and is reported in figure \ref{fig:PotTrack} for an external voltage of $V=25$~kV.

The precise evaluation of the PEs trajectories, the deposited energy in the scintillator and the SiPM light response is performed with GEANT4. For this study the software is instructed with the parameters of the SiPM selected for the characterization \cite{nuvfbk} (see section \ref{Sec:results}). PEs can be classified according to the trajectories. The first class is populated by straight PEs, that release the entire energy in the scintillator with a single interaction. As discussed in section \ref{Sec:abalone}, PEs can be scattered back, forming two more classes: returning PEs, if they release all the energy in the scintillator, and non-returning PEs, if the energy is not completely deposited in the windowlet. The last class of events is populated by undetected PEs.

\begin{figure}[htbp]
 \centering
 \includegraphics[width=0.5\textwidth]{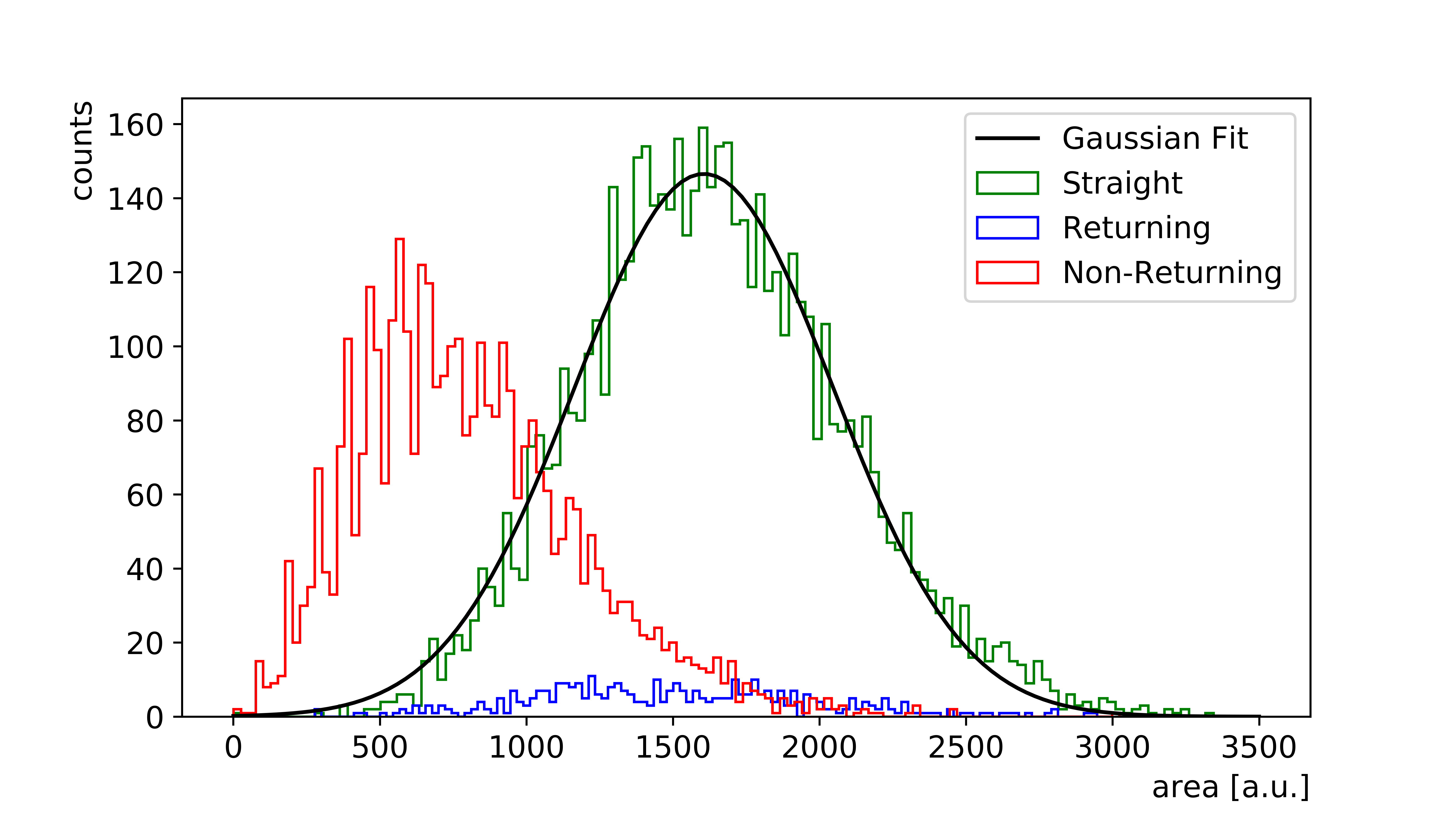}~
 \includegraphics[width=0.5\textwidth]{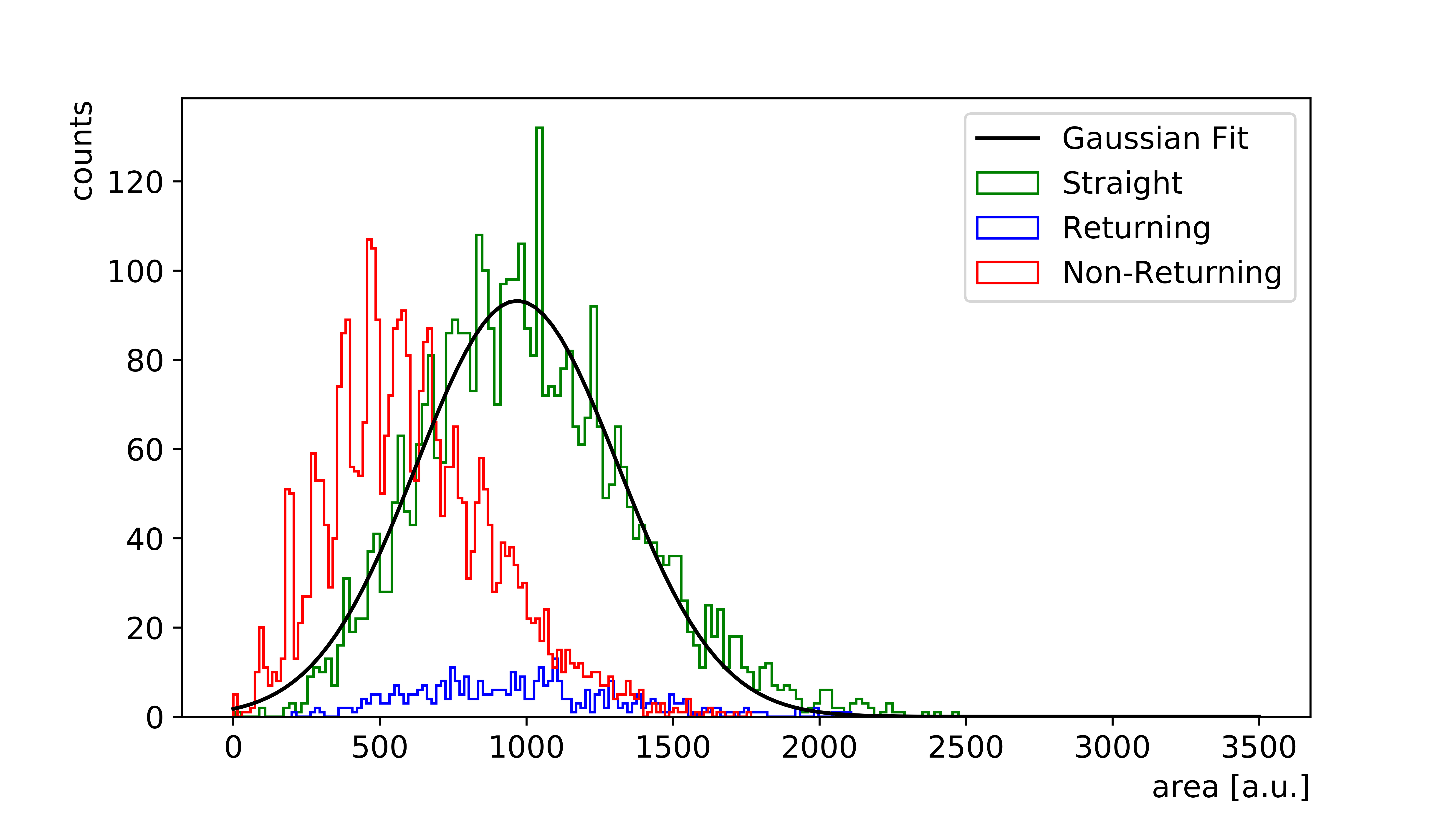}
 \caption{\label{Fig:spectra} Spectra obtained with simulation of PEs with an angle of 0 (left) and 83 (right) degrees. The different colors report the classes: straight (green), returning (blue) and non-returning (red) PEs.}
\end{figure}

Figure \ref{Fig:spectra} shows two example spectra obtained with PE simulations with an angle $\phi$, with respect to the ABALONE vertical axis, of 0 (left) and 85 (right) degrees. The colors distinguish the different classes. As expected straight and returning PEs produce Gaussian spectra, while the non-returning PEs spectrum exhibits a lower non-gaussian peak. 

\begin{figure}[htbp]
\begin{centering}
\includegraphics[width=0.5\textwidth]{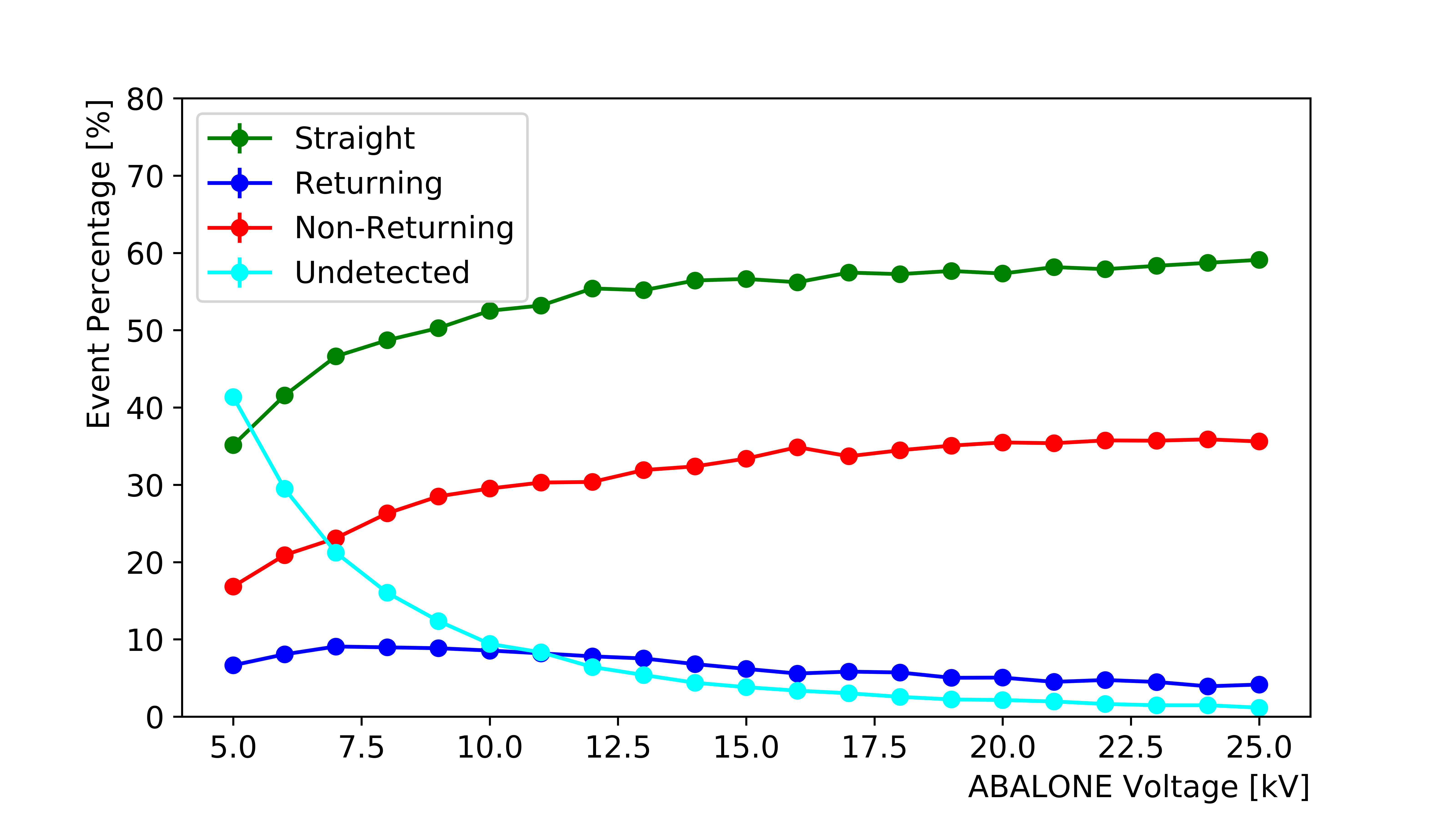}~
\includegraphics[width=0.5\textwidth]{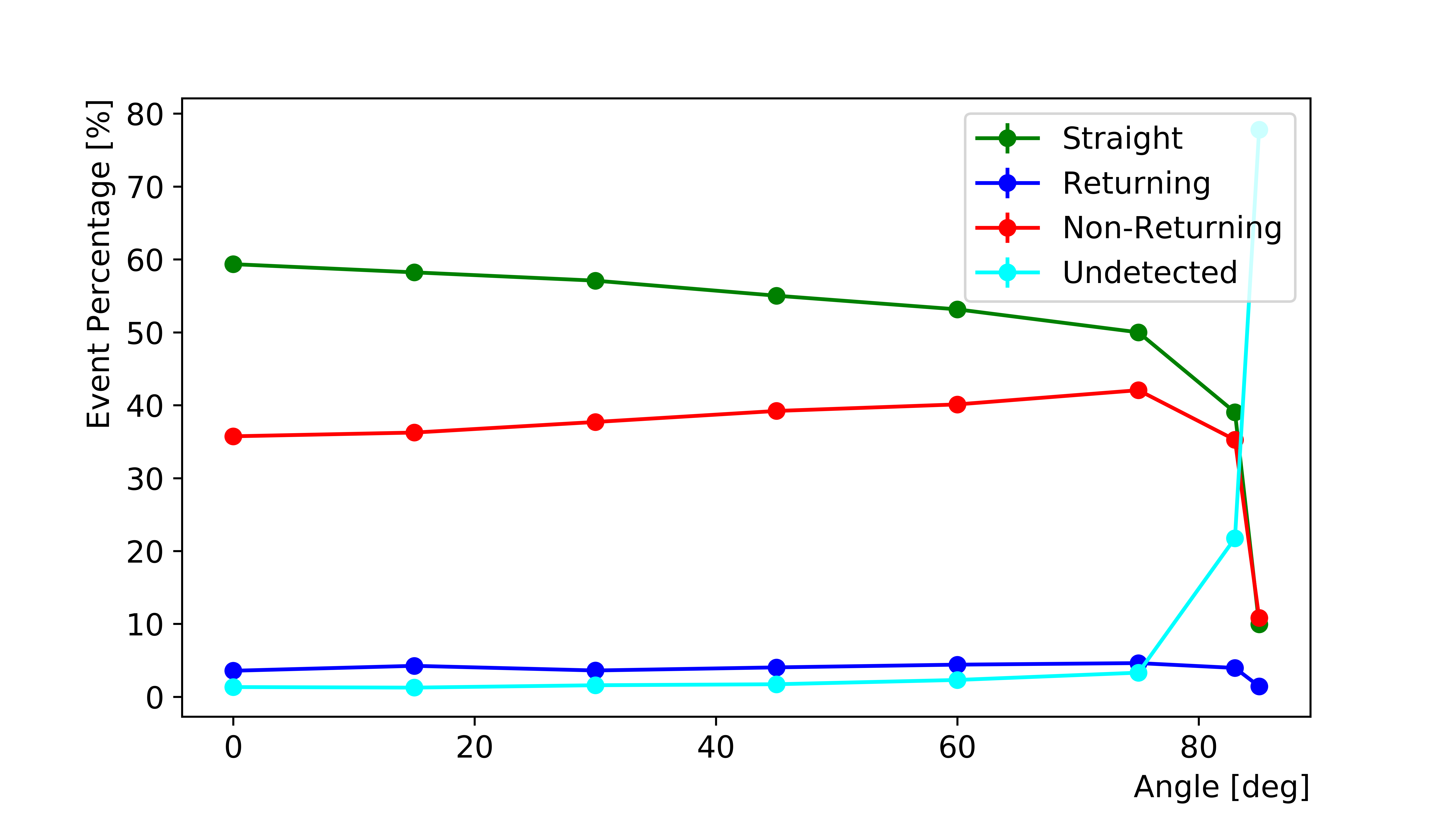}
\caption{Fraction of different PEs classes as function of voltage (left) and vertical angle (right).}
\label{Fig:fraction}
\end{centering}
\end{figure}

The distribution of the PE belonging to the three different classes have been studied by varying the ABALONE voltage between 5 and 25~kV, with $\phi=0~$degrees, and by simulating PEs emerging from vertical angle between 0 and 85 degrees, with a fixed $V=25$~kV. The results of this study are reported in figure \ref{Fig:fraction}: a voltage larger than 15~kV is needed to have a negligible fraction of undetected PEs (left plot), the photosensor is showing a good efficiency up to angles greater than 80 degrees (right plot). The fraction of non-returning PEs is around 35\% for $V>15~$kV and $\phi<80$~degrees, limiting the ABALONE detection efficiency.

\section{ABALONE characterization at room temperature}\label{Sec:results}
\subsection{Experimental Setup}

A test facility for ABALONE photosensor is realized at the Laboratori Nazionali del Gran Sasso. The ABALONE is set on a holder with the dome faced downward, the scintillation windowlet is optically coupled to the SiPM and a steel stand with a 420~nm LED pointing towards the dome, all inserted in a black box. The SiPM is a NUV-HD from FBK, a high-density SiPM optimized for positron emission tomography application, with photon detection efficiency of 63\% at 420 nm, 35~$\mu$m cell size, dark count rate of 100~kHz/mm$^2$ and an intrinsic gain of $4\times10^6$ \cite{nuvfbk}. The SiPM is connected to a preamplifier, then the signal is carried out of the box with a dedicated cable connected to a 100~MHz digitizer.

\subsection{Data analysis and results}
%The former thus contributes to the single-photon electron peak in the signal amplitude spectrum. 

The ABALONE is operated a high voltage up to 25~kV. The voltage is increased while continuously monitoring the dark count rate and performing dedicated LED data acquisitions. The goals of the first characterization stage is the calculation of the total gain of the ABALONE-SiPM system and the evaluation of PE classes fraction.

\begin{figure}[htbp]
\begin{centering}
\includegraphics[width=0.49\textwidth]{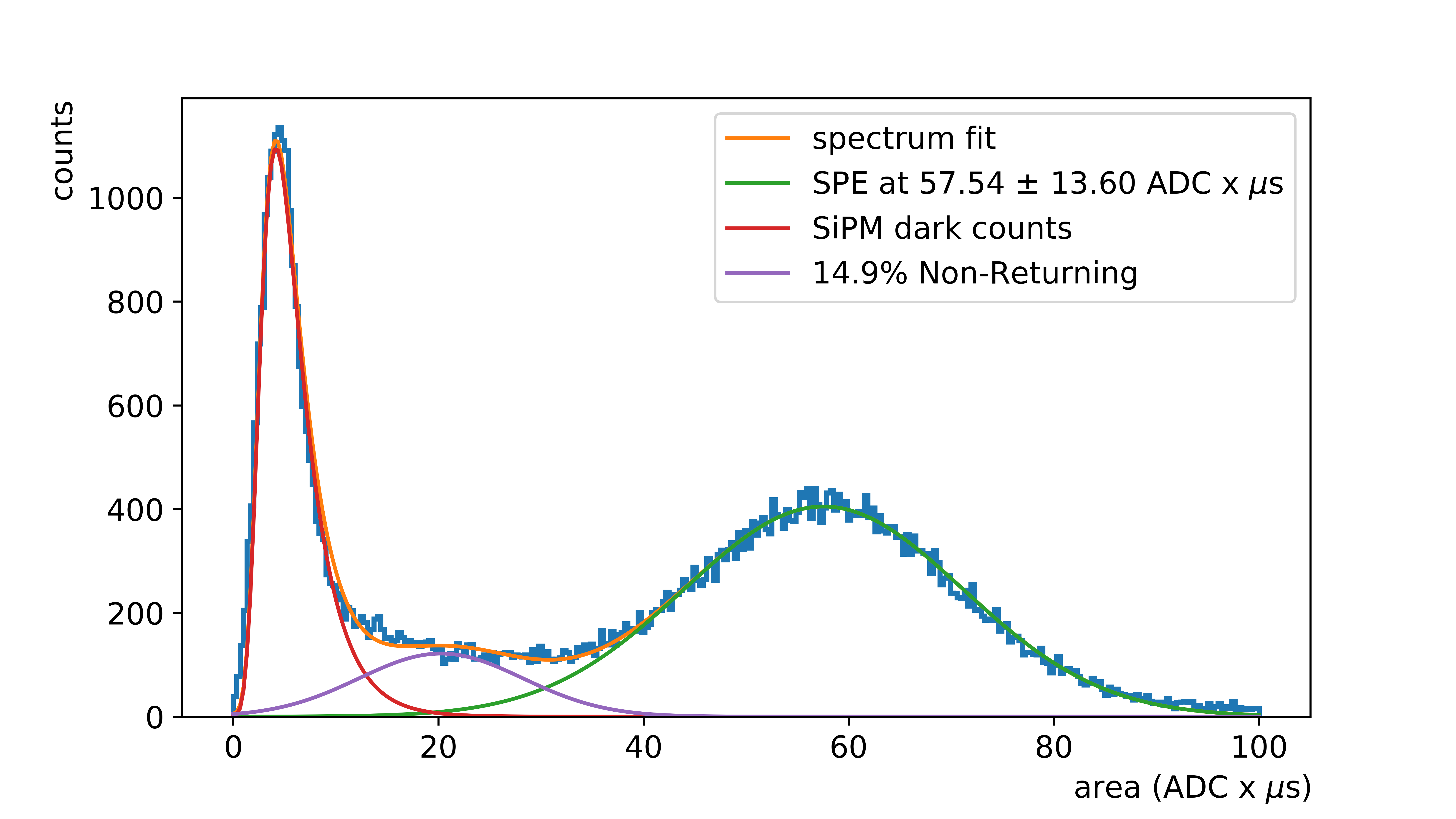}~
\includegraphics[width=0.49\textwidth]{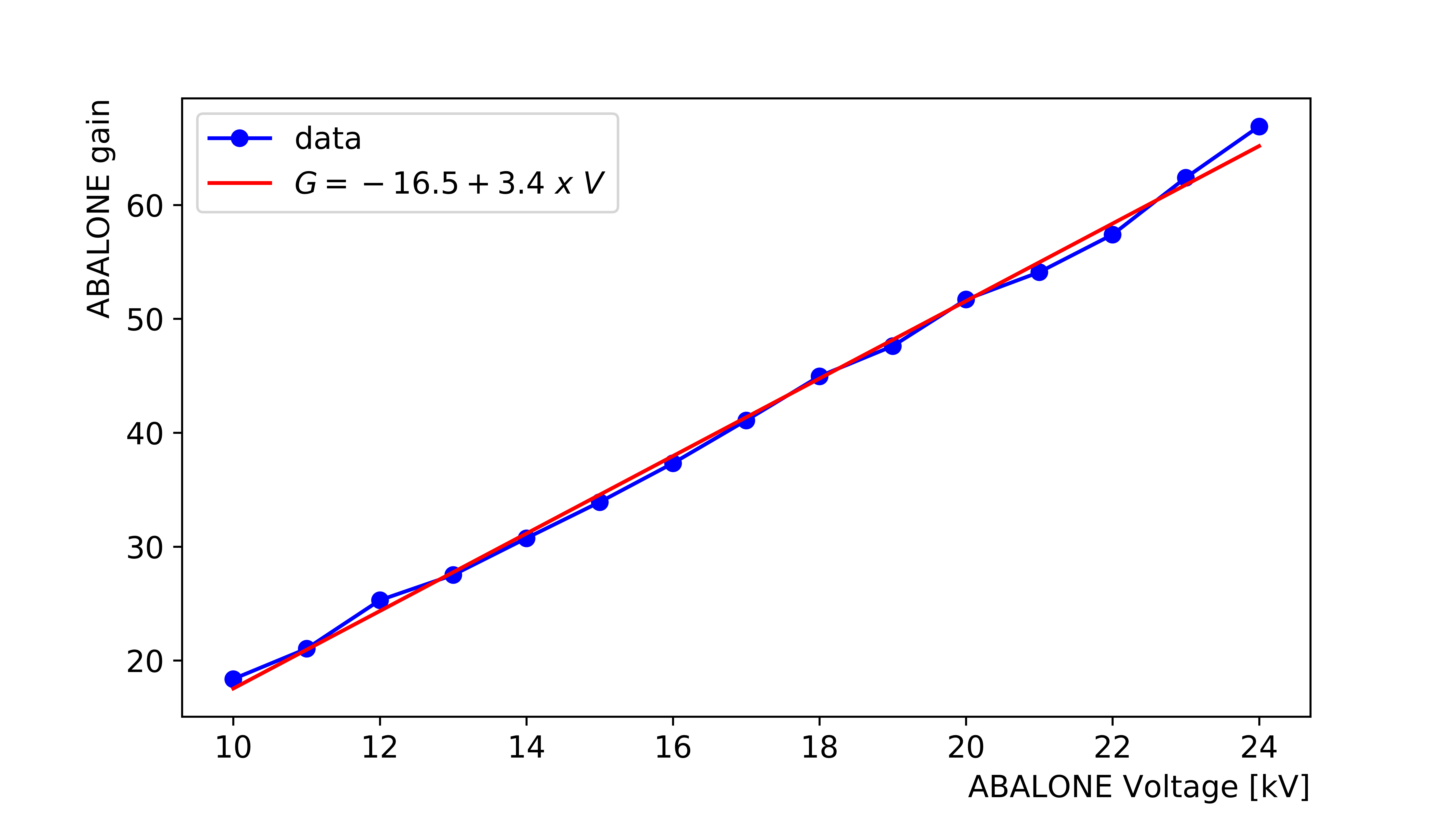}
\caption{Left: response peak area spectrum with $V=20$~kV. Right: evolution of the ABALONE gain as function of the applied high voltage.}
\label{Fig:results}
\end{centering}
\end{figure}

The gain of ABALONE is defined as the number of photons emitted by the scintillator for single photon received, while the SiPM gain is the number of PEs emitted for one detected photon. The gain of ABALONE-SiPM system and SiPM alone can be measured directly by looking at the area of the response peak. Figure \ref{Fig:results} (left) shows the spectrum for an ABALONE photosensor with $V=20$~kV: the first peak can be attributed to SiPM dark counts (following a Landau distribution), while the other peak is the single photoelectron (SPE) peak, resulting from the detection of straight and returning PEs. The region between the peaks can be attributed to non-returning PEs, since the area in this case is expected to span from zero to SPE peak. The fraction of non-returning PEs is around 15\% for a voltage higher than 15~kV. Anyway the distinction between the various PE classes is not trivial in experimental data and the current model needs to be optimized in future analysis.

The mean value of the SPE peak distribution (indicated as $\mu$) is exploited to calculate the ABALONE gain $G$, defined as:
\begin{equation}
    G = \frac{\mu_{\text{tot}}}{\mu_{\text{SiPM}}}
    \label{eq:gain}
\end{equation}
where $\mu_{tot}$ is extracted from the area spectrum of the complete system ABALONE-SiPM (e.g., 57.54~ADC$\mu$s in figure \ref{Fig:results}, left) and $\mu_{\text{SiPM}}$ is extracted from the area spectrum of SiPM alone. The results from the ABALONE gain calculated with voltage up to 24~kV is reported in figure \ref{Fig:results} (right): the data shows a linear trend $G=-16.5+3.4 \times V$ with a gain of 64.4 at 24~kV. Combined with the NUV-HD SiPM gain of $5\times 10^6$, they result in a combined gain of $3.2\times 10^8$, demonstrating the expected light intensification of the ABALONE.

\section{Conclusions and Outlook}

The ABALONE is an excellent candidate as replacement for PMTs in many fields of application, including large particle physics experiments and medical imaging, thanks to the cost effective, robust and performing technology. At the moment it is proposed as photosensor for the future dark matter experiment DARWIN \cite{darwin}. In this context the characterization of the ABALONE is being carried out at the Laboratori Nazionali del Gran Sasso and first results confirmed the main features of the photosensor and the expected light amplification. Future efforts aims to improve the ABALONE detection efficiency, in particular reducing the effect of non-returning PEs, exploiting the results from simulation studies. In collaboration with the University of Stockholm, in the near future the ABALONE will be tested in cold gas Xe and liquid Xe inside a cryogenic testing facility.

%Please also see the use of ``\texttt{\textbackslash texorpdfstring\{\}\{\}}'' to avoid warnings
%from the hyperref package when you have math in the section titles

%\acknowledgments

%The authors thanks the PhotonLab, Inc. and the directors and the staff of the Laboratori Nazionali del Gran Sasso for their support.

%\paragraph{Note added.} This is also a good position for notes added
%after the paper has been written.


\end{document}